\documentclass[acmsmall]{acmart}

\usepackage{booktabs}
\usepackage{tikz}
\usetikzlibrary{external}
\usepackage{ textcomp }
\usepackage[utf8]{inputenc}
\usepackage{enumitem}

\usepackage{epstopdf}
\usepackage{makecell}
\usepackage{threeparttable}
\usepackage{longtable}
\usepackage{multirow}
\usepackage{tabularx}
\usepackage{tabulary}
\usepackage{tablefootnote}
\usepackage{array}
\newcommand{\PreserveBackslash}[1]{\let\temp=\\#1\let\\=\temp}

\newcolumntype{C}[1]{>{\PreserveBackslash\centering}p{#1}}
\newcolumntype{R}[1]{>{\PreserveBackslash\raggedleft}p{#1}}
\newcolumntype{L}[1]{>{\PreserveBackslash\raggedright}p{#1}}
\usepackage{dsfont}
\usepackage{epstopdf}
\usepackage{graphicx}
\usepackage{subfig}
\usepackage{caption}
\usepackage{color}
\usepackage{mathtools}
\usepackage{url}
\usepackage{hyperref}
\usepackage{bm}
\usepackage{algorithmic}
\usepackage[linesnumbered,ruled,vlined]{algorithm2e}

\SetCommentSty{mycommfont}

\begin{document}
%\fancyhead{}

\title{Feature-level Attentive ICF for Recommendation}
\author{Zhiyong Cheng}
\affiliation{%
  \institution{Qilu University of Technology (Shandong Academy of Sciences)}
  \streetaddress{Shandong Artificial Intelligence Institute, 19 Keyuan Road}
  \city{Jinan}
  \state{Shandong}
  \postcode{250014}}
\email{jason.zy.cheng@gmail.com}

\author{Fan Liu}
\affiliation{%
  \institution{ Shandong University}
  \streetaddress{School of Computer Science and Technology, 72 Binhai Road}
  \city{Qingdao}
  \state{Shandong}
  \postcode{266200}
    \country{China}
  }
 \email{liufancs@gmail.com}

\author{Shenghan Mei}
\affiliation{%
  \institution{ Shandong University}
  \streetaddress{School of Computer Science and Technology, 72 Binhai Road}
  \city{Qingdao}
  \state{Shandong}
  \postcode{266200}
    \country{China}
  }
 \email{hgsqtdsswk@gmail.com}

\author{Yangyang Guo}
\affiliation{%
  \institution{ Shandong University}
  \streetaddress{School of Computer Science and Technology, 72 Binhai Road}
  \city{Qingdao}
  \state{Shandong}
  \postcode{266200}
   \country{China}
 }
\email{guoyang.eric@gmail.com}

\author{Lei Zhu}
\affiliation{
    \institution{Shandong Normal University}
    \streetaddress{College of Computer Science and Electronic Engineering}
    \city{Jinan}
    \state{Shandong}
    \postcode{250358}
    \country{China}
}
\email{leizhu0608@gmail.com }

\author{Liqiang Nie}
\affiliation{%
  \institution{ Shandong University}
  \streetaddress{School of Computer Science and Technology, 72 Binhai Road}
  \city{Qingdao}
  \state{Shandong}
  \postcode{266200}
  \country{China}
}
\email{nieliqiang@gmail.com}

\begin{abstract}
Item-based collaborative filtering (ICF) enjoys the advantages of high recommendation accuracy and  ease in online penalization and thus is favored by the industrial recommender systems. ICF recommends items to a target user based on their similarities to the previously interacted items of the user. Great progresses have been achieved for ICF in recent years by applying advanced machine learning techniques (e.g., deep neural networks)  to learn the item similarity from data. The early methods simply treat all the historical items equally and recently proposed methods attempt to distinguish the different importance of historical items when recommending a target item. Despite the progress, we argue that those ICF models neglect the diverse intents of users on adopting items (e.g., watching a movie because of the director, leading actors, or the visual effects). As a result, they fail to estimate the item similarity on a finer-grained level to predict the user's preference to an item, resulting in sub-optimal recommendation. In this work, we propose a general feature-level attention method for ICF models. The key of our method is to distinguish the importance of different factors when computing the item similarity for a prediction. To demonstrate the effectiveness of our method, we design a light attention neural network to integrate both item-level and feature-level attention for neural ICF models. It is model-agnostic and easy-to-implement. We apply it to two baseline ICF models and evaluate its effectiveness on six public datasets. Extensive experiments show the feature-level attention enhanced models consistently outperform their counterparts, demonstrating the potential of differentiating user intents on the feature-level for ICF recommendation models.

\end{abstract}

\begin{CCSXML}
<ccs2012>
    <concept>
        <concept_id>10002951.10003317.10003331.10003271</concept_id>
        <concept_desc>Information systems~Personalization</concept_desc>
        <concept_significance>500</concept_significance>
        </concept>
        <concept>
        <concept_id>10002951.10003317.10003347.10003350</concept_id>
        <concept_desc>Information systems~Recommender systems</concept_desc>
        <concept_significance>500</concept_significance>
        </concept>
        <concept>
        <concept_id>10002951.10003227.10003351.10003269</concept_id>
        <concept_desc>Information systems~Collaborative filtering</concept_desc>
        <concept_significance>500</concept_significance>
    </concept>
</ccs2012>
\end{CCSXML}

\ccsdesc[500]{Information systems~Personalization}
\ccsdesc[500]{Information systems~Recommender systems}
\ccsdesc[500]{Information systems~Collaborative filtering}

\keywords{Attention, Collaborative Filtering, Diverse preference, Item-based Recommendation}
%%%%%%%%%%%%%%%%%%%%%%%%%%%%%%%%%%%%%%%%%%%%%%%%%%%%%%%%%%%%%%%%%%%%%%%%%%%%%%%%

\maketitle

\section{Introduction}
In the information age, we face overwhelming information at almost all aspects of our work and life.  How to quickly find the desired information has thus become crucial in our daily lives. Recommendation as an effective information filtering and seeking technique~\cite{Koren2009Matrix, He2017Neural,He2020lightgcn} has been widely deployed in current online service platforms, including information/media provider, E-commerce and social platforms.
Among various recommendation methods, collaborative filtering (CF)~\cite{Jonathan2017Algorithmic,Sarwar2001Item,Zhang2016Discrete,liu2021interest} is one of the most dominant recommendation techniques and has attracted a lot of attention from researchers and practitioners since its birth. In general, CF methods can be categorized into two paradigms: user-based CF (UCF) and item-based CF (ICF)~\cite{su2009survey}. The key of UCF is that users sharing close preferences often like the same items. That is, previously consumed items by one user will be recommended to another similar user with a large probability. In contrast, ICF methods represent a user with all his/her historically consumed items~\cite{Sarwar2001Item}. Specifically, the similarities between the target item and the previously interacted items are estimated firstly, which are then treated as the pivot for recommending similar items to the target user.

Comparing to UCF, ICF enjoys several advantages in practice. Firstly, ICF models a user preference based on her previously interacted items, which enables it encode more signal in its input than UCF that simply uses an ID embedding~\cite{He2018Nais,Xue2019Deep}.  The user preference on items  are relatively stable unless the background (or context) has dramatically changed, especially for the long-term preference. The aspects (or characteristics) that a user cares in the past will be also important for her in a long time. The ICF models represent a user by previous interaction items, which is actually profiling the user with the characteristics of those interacted items. This empowers ICF more potential to improve the recommendation performance~\cite{Kabbur2013Fism,christakopoulou2016local,He2018Nais,Xue2019Deep}. Secondly, ICF has better interpretation, because it can explain a recommendation with similar items that the user interacted before. This is more acceptable for  users than the ``similar users" based explanation, as those similar users might be strangers for the target user.
In addition, ICF is flexible to incorporate new user-item interactions into the model, which makes it more suitable for online personalization~\cite{He2018Nais,Xue2019Deep}. For new interactions, UCF methods need to re-train the model for updating the user representations, which is very time-consuming and impractical in industrial applications. On the contrary, ICF can simply retrieve items similar to the newly interacted ones (i.e., leveraging item similarities) and recommend them to the current user. It does not need the model re-training processing and thus is more time efficient~\cite{Covington2016Deep,Eksombatchai2018Pixie,Smith2017Two}.

Early ICF approaches estimate the item similarities using statistical measures, such as Pearson coefficient~\cite{Koren2008Factorization} or cosine similarity~\cite{Sarwar2001Item}. The main drawback of  those heuristic approaches is that they often require heavy manual tuning on the similarity measure for good performance on a target dataset. As a result, such methods are hard to be directly applied to a new dataset. To tackle this limitation, data-driven methods~\cite{Kabbur2013Fism,Ning2011Slim} have been developed to learn item similarity from data. These methods first calculate the final result by setting an objective function, and then calculate the parameters by passing data into the loss function. Theoretically, the richer the data, the more accurate the model can be. In addition, the data-driven methods save the time of parameter adjustment. They can not only improve the efficiency but also enjoy higher accuracy,  because the calculation of the parameters is based on the real data and does not rely on the experience of the participators. Recently, He et al.~\cite{He2018Nais} pointed out that existing data-driven ICF methods assume all historical items of a user contribute equally in estimating the target item for the target user,  resulting in sub-optimal performance. They therefore developed a neural attentive item similarity model (NAIS) to distinguish the different importance of previously interacted items for the user preference to the target item.

Though NAIS has achieved superior performance over existing ICF methods, we argue that its performance is still limited because it neglects users' diverse intents on adopting items. More specifically, a user often pays attention to certain features when selecting an item to consume. Accordingly, those features will dominate the attitude of user preference towards this item. In addition, for each user, the dominant features are usually different from item to item. For example,  a user may favor a movie because of its plot, and likes another movie because she is a fan of its director. With this consideration, we deem that treating all the features\footnote{In this paper, we regard different dimensions of the item embedding as different features that reflect user intents.} equally is not optimal in recommendation. However, it is not straightforward to explicitly model the  impact of different features in ICF for recommendation, because ICF models rely on estimating the similarity between the target item and historical items for prediction.  NAIS~\cite{He2018Nais} has demonstrated the importance of  item-level attention  for ICF recommendation methods. To further enhance the recommendation accuracy, it is necessary to consider both item-level and feature-level attentions simultaneously in the model. How to combine them without complicating the model and increasing the computational burden much is another problem to be solved.  

In this work, we make an effort to tackle aforementioned problems and present a general feature-level attention framework for ICF models to consider users' diverse intents in recommendation. Our proposed method models a user's diverse intents by distinguishing the impact of different features of a historical item to the target item in prediction. Concretely, our method computes a weight vector for each historical item to estimate the similarity between this historical item and the target item. This weight vector is used to differentiate the contributions of different features in prediction by assigning different weights to each feature of the embedding vector.  Based on this idea, we further design an attention neural network to combine the item- and feature-level attention for neural ICF models. It is light and easy-to-implement in different ICF models. To evaluate its effectiveness, we apply it to two models NAIS~\cite{He2018Nais} \& DeepICF~\cite{Xue2019Deep} and  conduct experiments on six Amazon datasets. Extensive experimental results show that the feature-level attention enhanced models can indeed improve the performance consistently  over their counterparts (i.e., NAIS and DeepICF) and achieve the state-of-the-art performance. 

In summary, the main contributions of this work  are threefold:
\begin{quote}
    \begin{itemize}
        \item We highlight the importance of considering users' diverse intents in ICF methods and propose to model the intents on the feature level. In particular, we present a general  feature-level attention (FLA) method to measure the importance of different features of a historical item to the target item in ICF models.
        \item We design a light and model-agnostic attention neural network which can effectively combine the item- and feature-level attention for neural ICF models. It is easy to implement in neural ICF models and we apply it to the NAIS and DeepICF to enhance their performance.
        \item We conduct extensive experiments on six publicly available datasets and demonstrate the effectiveness of our proposed method. Experimental results show that the FLA-enhanced NAIS and DeepICF can achieve better performance, demonstrating the effectiveness of the proposed method. We released our codes and the parameter settings for the experiments to facilitate others to repeat this work.\footnote{https://github.com/liufancs/FLA}
    \end{itemize}
\end{quote}

The rest of this paper is organized as follows. We first review related work  in Section~\ref{sec2}. In Section~\ref{sec4}, we elaborate our feature-level attention method  and then describe its application to NAIS and DeepICF in Section~\ref{sec:app}. In the next, we report the experimental results in Section~\ref{sec5}. Finally, we conclude the paper in Section~\ref{sec6}. 

\section{Related Work} \label{sec2}
\subsection{Collaborative Filtering}
Collaborative filtering (CF)~\cite{Koren2010Factor,Hu2008Collaborative,Pan2008One} has long been recognized as an effective approach in recommendation over the past decades. Based on the standpoint of the interacted instances, CF methods can be classified into two categories: user-based CF (UCF) and item-based CF (ICF). The former one recommends a user with the items favored by her similar users; and the latter one recommends a user with the items that are similar to the items she liked in the history. UCF has been extensively studied in both academia and industry. A typical UCF method is matrix factorization (MF)~\cite{Koren2009Matrix}, which represents users and items as
feature vectors in the same embedding space based on the user-item interactions, and then predicts the preference of a user to an item by an interaction function (i.e., inner product) between their
embedding vectors. This simple idea has achieved great success in the Netflix contest and many variants have been developed later on, such as WRMF~\cite{Hu2008Collaborative}, BPR~\cite{Rendle2009Bpr}, NeuMF~\cite{He2017Neural}, LightGCN~\cite{lightgcn}. Although UCF has achieved significant progress, a big limitation is that the UCF models require to be re-trained when new interactions come in, which is unacceptable in real-time recommender systems~\cite{He2018Nais,Xue2019Deep}. In contrast, ICF predicts user preference to a target item by estimating the similarity scores between the previously interacted items of this user and the target one, which enables ICF to easily incorporate new interactions into the preference modeling. Due to the nice property of ease online updating, ICF models are favored by industry and have been widely-adopted in real recommender systems~\cite{Deshpande2004Item,Guo2019Locally,Wang2019Neural}.

Early ICF models leverage heuristic metrics, such as cosine similarity~\cite{Sarwar2001Item} or Pearson correlation coefficient~\cite{Koren2008Factorization} to calculate the similarity, which require quite a lot of manual tuning when adapting to another brand-new dataset. In order to tackle this limitation, several data-driven methods have been proposed~\cite{Wu2016Collaborative,Christakopoulou2018Local}. For example, SLIM~\cite{Ning2011Slim} learns a complete item-item similarity matrix by minimizing the errors between the reconstructed rating matrix and the ground-truth. With the designed object function optimized for recommendation, SLIM can achieve higher
recommendation accuracy. Christakopoulou et al.~\cite{christakopoulou2016local}  extended SLIM to model the preference of like-minded usersets and proposed  a global and local SLIM (GLSLIM) method . This model applies different SLIM models to capture the preference of different user subsets.  Later on, they  pointed out that high-order item relations also provide valuable information for user preference modeling and proposed a higher-order sparse linear method (HOSLIM)~\cite{Christakopoulou2014Hoslim}, which extends the SLIM model to learn the item-itemset similarity for capturing the higher-order relations. More recently, Xue et al. ~\cite{Xue2019Deep} proposed a DeepICF model, which captures the higher-order item relations by stacking multiple layers over the second-order item relations in a non-linear way. 

A big limitation of SLIM is that it's very space- and time-consuming to learn the similarity matrix, which makes it unscalable and limits its application in real systems, considering the tens of millions of items in modern E-commerce platforms. Besides, the transductive relations are omitted since only co-interacted items are considered. To address the limitation, FISM~\cite{Kabbur2013Fism} first represents each item as a low-dimensional vector and then models the similarity between each pair of items by the inner product of their embedding vectors.   He et al.~\cite{He2018Nais} pointed out that FISM treats each item equally for
the preference prediction to the target item. However, this is often not true in practice, because some items are more relevant to the target item. With this consideration,   they  developed an attention-based method called NAIS~\cite{He2018Nais} to assign different weights to the historical items for better capturing user preference. %An early neural network based ICF model is the CADE model~\cite{Wu2016Collaborative}, which learns the item similarity by using nonlinear auto-encoder architecture. 

Despite great progress has been achieved by those ICF models, those models have not considered user diverse intents towards different items in an explicitly way. 
In this paper, we make an effort to model user preference at the feature-level (i.e., each feature is considered as an intent dimension) in the ICF model and propose a feature-level attention method to enhance the performance of ICF models.

\subsection{Attention-based Recommendation}
The attention mechanism has been widely-used in deep learning methods and achieved great success in many tasks in computer vision and natural language processing. With the widespread application of deep learning in recommendation, this technique has also been used in various ways in recommender systems in order to model user preference more accurately. Many attention-based recommender systems have been developed. A comprehensive survey of attention-based recommender system is out-of-the-scope of this paper. In this section, we briefly review the two paradigms of using attention-mechanism in recommender systems.  

\textbf{Item-level attention.} As discussed, historically interacted items have different contributions to model users' preference. Therefore,  it is important to assign different weights to the items for more accurate recommendation ~\cite{Ebesu2018Collaborative,Zheng2019Mars,Zhou2018Deep}.  NAIS~\cite{He2018Nais} and DeepICF~\cite{Xue2019Deep} are typical examples of this paradigm. Besides the ICF models, the item-level attention has also been used in graph convolution network (GCN) based recommender systems. The core of GCN-based recommendation models is that the embeddings of users/items are iteratively updated by aggregating information from their local neighbors (i.e., interacted items/users)~\cite{pinsage,ndcg2002,He2020lightgcn}. The attention mechanism is introduced to differentiate the different contributions of neighboring nodes in the user/item embedding learning process~\cite{wei2019hashtag,liu2020tkde, xiao2020MCCF}.  Another widely applied task for item-level attention is the the session-based recommendation task. Because interacted items in a session are typically sparse, it becomes very crucial to identify important items for user intent inference~\cite{session2020survey}. A general framework is to use a recurrent neural network to learn the hidden states of items inside a session, followed by an attention model on the items' hidden representations to capture the main purpose of users~\cite{Tan2016Improved,Li2017Neural}. Recently, the self-attention blocks, such as Transformer~\cite{Vaswani2017Attention} and BERT~\cite{Devlin2019Bert} have also been applied to the session-based recommendation~\cite{Kang2018Self,Anh2019Session}.  

\textbf{Feature-level attention in side information.}  The attention mechanism has become a standard component in the side information enriched recommender system, in order to extract effective features from the side information to represent item features or user preference. The most widely used side information is review and user/item attributes. At the beginning,  the attention mechanism is only used to assign different weights on the review-level for learning user and item embeddings~\cite{chen2018www,liu2020neural,wu2019tois}. A representative method is A$^3$NCF~\cite{Cheng2018A3ncf}. In this method, for each user-item pair, it learns the attentive weights for each factor by taking the user's and item's embedding, as well as their text-based representations learned from review into an attentive neural network.  Later on, the review-aware recommender systems exploit the reviews at a more fine-grained level by applying the attention mechanism in a hierarchical manner~\cite{cong2019hierarchical,liu2019npra}: 1) first attending important words of a review (i.e., word-level) to learn better review representations, and then 2) assigning different weights to review representations for user and item embedding learning. Beyond the two-layer of hierarchical attention network design, Wu et al.~\cite{wu2019hierarchical} proposed to additionally encode the sentence-level attention in the review and developed a three-tier attention network for recommendation.  Besides, there are also aspect-aware attention-based recommendation models~\cite{guan2019tois,Chin2018Anr}, which extract aspects from the reviews and then assign weights to different aspects in the user preference modeling.

Attribute information is often used in factorization machine~\cite{rendle2010factorization} and graph-based models~\cite{liu2020tkde}, especially knowledge-graph (KG) based recommendation models~\cite{kgat,hu2018kdd,wang2019tois,shi2019tkde,kgrec2020survey}. A representative attention-based FM based model is the AFM model~\cite{xiao2017afm}, which learns the importance of each feature interactions from data via a neural attention network. In the KG-based recommender systems, the attributes of items/users are taken as node entities in the graph. There are two typical ways of applying KGs: embedding-based and meta-path based. In the embedding-based models, such as KGAT~\cite{kgat}, RippleNet~\cite{wang2019tois}, AKGE~\cite{sha2019attentive}, and A$^2$-GCN~\cite{liu2020tkde}, the attention mechanism is often used to learn the importance of neighbor nodes during the embedding propagation. For meta-path approaches, the attention mechanism can be applied inside a meta-path to learn representation of the meta-paths or directly attends to different meta-paths. A typical meta-path based recommendation approach is MCRec~\cite{hu2018kdd}, which first uses the attention mechanism to learn the representation of meta-paths and then applies it to assign the weights of different meta-paths for final user representation learning. 

The attention mechanism is also used in visual-aware and multimedia  recommendation. For example, Chen et al.~\cite{Chen2017Attentive} proposed an ACF model for multimedia recommendation, in which a component-level attention model is used to capture the user's different preferences on different components, e.g., certain actions in a video; and an item-level attention model is leveraged to treat historically interacted items differently. In~\cite{chen2019visual}, a visually explainable recommendation model is presented to capture use attention on different regions of images based on attention neural networks.

%\textbf{Factor-level attention.} In this approach, the attention weights are assigned to different factors of the target item embedding to capture the user's specific preference on this item. From this perspective, a representative method is A$^3$NCF~\cite{Cheng2018A3ncf}. In this method, for each user-item pair, it learns the attentive weights for each factor by taking the user's and item's embedding, as well as their text-based representations learned from review into an attentive neural network. Note that there is a big difference between the A$^3$NCF and the method presented in this work. The A$^3$NCF is a user-based CF model which learns the attentive weights based on the target user and item embeddings; and our method here is designed for item-based CF methods, which do not modeling user embeddings.

\subsection{Diverse Preference Modeling}
The underlying rationale of our feature-level attention is that a user's intent to different items could be diverse. Traditional recommender systems often represent a user preference with a fix embedding vector, which is then used to match the vectors of different items for preference prediction. This process does not differentiate user intents on different items. In recent few years, researchers start to pay attention to model the diverse preferences of users towards different items and proposed several methods. Cheng et al.~\cite{cheng2018aspect,cheng2019tois,Cheng2018A3ncf} proposed to model user intents on different aspects of items.  They first applied topic models on side information (e.g., reviews and images) to analyze user interests on different aspects of items. These aspects are then linked to the factors of (user/item's) embeddings (learned by matrix factorization~\cite{cheng2018aspect,cheng2019tois} or neural networks~\cite{Cheng2018A3ncf}). For a target user-item pair, a unique weight vector is learned to represent this user's attention on different factors of the target item. This unique weight vector is expected to capture the user's intent (e.g., on which aspects) to the target item. Following this idea, Chin et al.~\cite{Chin2018Anr} presented an end-to-end neural recommendation model called ANR, which exploits the review information to model user diverse preference on different aspects of items. Later on, Liu et al.~\cite{liu2019mm} presented a metric learning based recommendation model, which uses an attentive neural network to estimate user attention on different aspects of the target item by exploiting the item's multimodal features (e.g., review and image). 

To take the user diverse preference on items into consideration, another line of work is to dynamically adapt the target user's or item's embedding to accurately predict the user preference to the target item. For example, CMN~\cite{Ebesu2018Collaborative} adapts the target user embedding based on the selected most influential neighbor users, whose influential scores are computed according to the target item. MARS~\cite{Zheng2019Mars} adopts a different strategy, which adapts the user vector embedding based on the most influential item vectors of the target item. In contrast, DIN~\cite{Zhou2018Deep} adapts the target item embedding based on the user's previously purchased items. More recently, the disentangled representation learning approach has been applied in recommendation for disentangled embedding learning.  A representative method is the disentangled graph collaborative filtering (DGCF) method proposed by Wang et al.~\cite{wang2020sigir}. In this method, different intents are represented as different chunks in the embedding vector and  a distance correlation regularization is applied to make those chunked representations independent. Different from this method, DisenHAN~\cite{disenHAN} learns the disentangled representations by aggregating aspect features from different meta relations in a heterogeneous information network (HIN). 

Apparently, the method presented in this work is fundamentally different from the above method. Our method predicts user preference on the target item by attending each feature in the item embedding vectors of the historical items.  All the above methods fall into the user-based CF approach, and they use the learned user embedding to analyze the user intent to the target item.

\section{Feature-level Item Attention} \label{sec4}
The underlying intuition of NAIS~\cite{He2018Nais} is that the more relevant of a historical item to the target item, the more important role it plays in the preference prediction. Therefore, NAIS introduces an attention mechanism to estimate the contribution of each historical item to the target item.  The item-level attention only computes an attentive weight based on the overall relevance between the two items while ignores the attention on different features, and thus fails to capture the fine-grained user preference. It is well-known that an item is depicted by different features~\cite{Cheng2018A3ncf} and the preference of a user to an item often depends on a few features, such as the \emph{directors} or \emph{actors} of an movie. Therefore, a user $u$'s preference to an item $i$ depends on \emph{$u$'s attention on which features of the item} and \emph{whether those features of the item fit the user's tastes}. Based on this consideration, when considering the contribution of a historical item to the target item in preference prediction in ICF, it is better to measure the importance of different features of the historical item. In this section, we will introduce a feature-level attention method, which considers the contribution of historical items on the feature-level for ICF recommendation. We first introduce the general method of computing the feature-level attention between two items, and then introduce the method to consider both item-level and feature-level attention in ICF models. %Finally, we show the application of our proposed method in the recently advanced NAIS and DeepICF models.

\subsection{Feature-level Attention} \label{sec:fla}
In the embedding-based ICF models (such as FISM~\cite{Kabbur2013Fism}, NAIS~\cite{He2018Nais}, DeepICF~\cite{Xue2019Deep}), items are mapped into a latent feature space and each item is represented by a vector in this space.  Let $\mathbf{p_i} \in  \mathds{R}^d$ and $\mathbf{q_j} \in  \mathds{R}^d$ denote the feature vector of the target item $i$ and a historical item $j$, respectively. $d$ is the dimensionality of the latent space and each dimension can be regarded as a certain feature to describe the items. Our intuition is that different features of a historical item $j$ contribute differently to a target item $i$. Taking a toy example: if we only use two features - ``leading actors" and ``director" to describe a movie, given a historical movie $m_0$ of a user $u$, it has the same \emph{leading actors} but different \emph{directors} from a movie $m_1$; and for another movie $m_2$, it has different \emph{leading actors} but the same \emph{director}. When predicting $u$'s preference to $m_1$, the feature of ``leading actors" should play a more important role than that of the ``director"; when predicting  $u$'s preference to $m_2$, the feature of ``director" will be more important. Therefore, for each historical item $j \in \mathcal{R}_u^{+}\backslash\{i\}$, our goal is to compute an attentive vector $\mathbf{a_{ij}}$, in which each element $\{a_{ijk}|k=\{1, \cdots, d\}\}$  indicates the importance of $k$-th feature of the item $j$ with respect to the target item $i$.

Notice that the value of $a_{ijk}$ indicates the relatively importance of the $k$-th feature, it depends on the similarity of other features between the two items $i$ and $j$. For example, if all the features of the two items are the same (e.g., the leading actors and directors are all the same for two movies), the features are equally important. Inspired by the effectiveness of NAIS and DeepICF, we first use the element-wise product on the vectors of two items to obtain an interacted vector, i.e., $\mathbf{p_i} \odot \mathbf{q_j}$, where $\odot$ indicates element-wise product.  We then follow the standard attention mechanism by applying a non-linear transformation to obtain the attentive weights:
\begin{equation} \label{eq:fla}
\mathbf{\hat{a}_{ij}}=\mathbf{H}^\top ReLU(\mathbf{W}(\mathbf{p}_i\odot\mathbf{q}_j)+\mathbf{b}) ,\\
\end{equation}
where $\mathbf{W} \in \mathds{R}^{d' \times d}$ and $\mathbf{b} \in \mathds{R}^{d'}$ denote the weight matrix and bias vector of the attention network, respectively. $\mathbf{H} \in \mathds{R}^{d' \times d}$ denotes the weight matrix of the output layer of the attention network.\footnote{Note that in the NAIS, it is a weight vector $\mathbf{h} \in \mathds{R}^{d'}$ for the output layer of the attention network.} $d'$ denotes the size of the hidden layer. Because our goal is to compute the importance of different features of the items. The softmax function is then used to normalize the attentive weights:
\begin{equation} \label{eq:softmax}
   a_{ijk}=\frac{\exp(\hat{a}_{ijk})}{\sum_{k'=1}^{d}\exp(\hat{a}_{ijk'})}.
\end{equation}

\textbf{Discussion.} In this design, the attentive weights of features are computed based on the interaction between the vectors of two items. Theoretically, other functions can be also applied to encode the interaction, for example,  addition, subtraction, etc. We use the element-wise product because it is a generalization of inner product to vector space. Notice that each element in  $\mathbf{v_{ij}}$ (i.e., $v_{ijk}$) is a product of the corresponding feature of the two vectors (i.e., $p_{ik} \cdot q_{jk}$), and it can be regarded as a similarity of the corresponding feature between two items. This is similar to the inner product to compute the similarity between two vectors.

The proposed feature-level attention looks similar to that of NAIS, because both methods compute the attention of a historical item to a target item. The difference is that NAIS computes an attentive weight for the historical item, but we compute an attentive weight vector for all the features of the historical item. Our method takes one-step further to consider the different contributions of features in items than NAIS which considers the contributions of different items. As aforementioned, our model computes the relatively importance of each feature among all the features of an item. Considering that historical items have different relevance levels to a target item, we should also consider the item-level attention simultaneously. Because for a target item, the features with high attentive weight of an irrelevant item could contribute less than the features with relatively low attentive weights of a relevant item. In the next section, we introduce our design to consider both item-level attention and feature-level attention in ICF models.

%The item-level model aims to assign different weights to all the historical items interacted by the target user, which neglects the different contribution of item features. However, these item features do affect user preferences to some extent. For example, the \emph{movie} item contains \emph{actors} and \emph{director} features. A user chooses to watch some \emph{movies} because all these movies are directed by the same \emph{director} and the user is a fanatical fan of the \emph{director}. Based on this assumption, we intuitively combine each item features in a more detailed perspective, and our designed method computes an attention vector instead of a value for each item. The intuition behind this is that we intend to measure the contribution of each item feature. The $\mathbf{a}_{ij}$ can be computed as,

%Under this design, we take the feature-level influences for user preference modeling into consideration. For instance, if a user prefers to watch movies directed by a certain director, the feature \emph{director} will affect more on the recommendation of the next expected \emph{movie}. However, different items also contribute discriminately to the user preferences, we thus add the item-level influence in the next.

\subsection{Item- and Feature-level Attention} \label{sec:ifla}
\begin{figure}[t]
    \centering
    \includegraphics[width=0.9\textwidth]{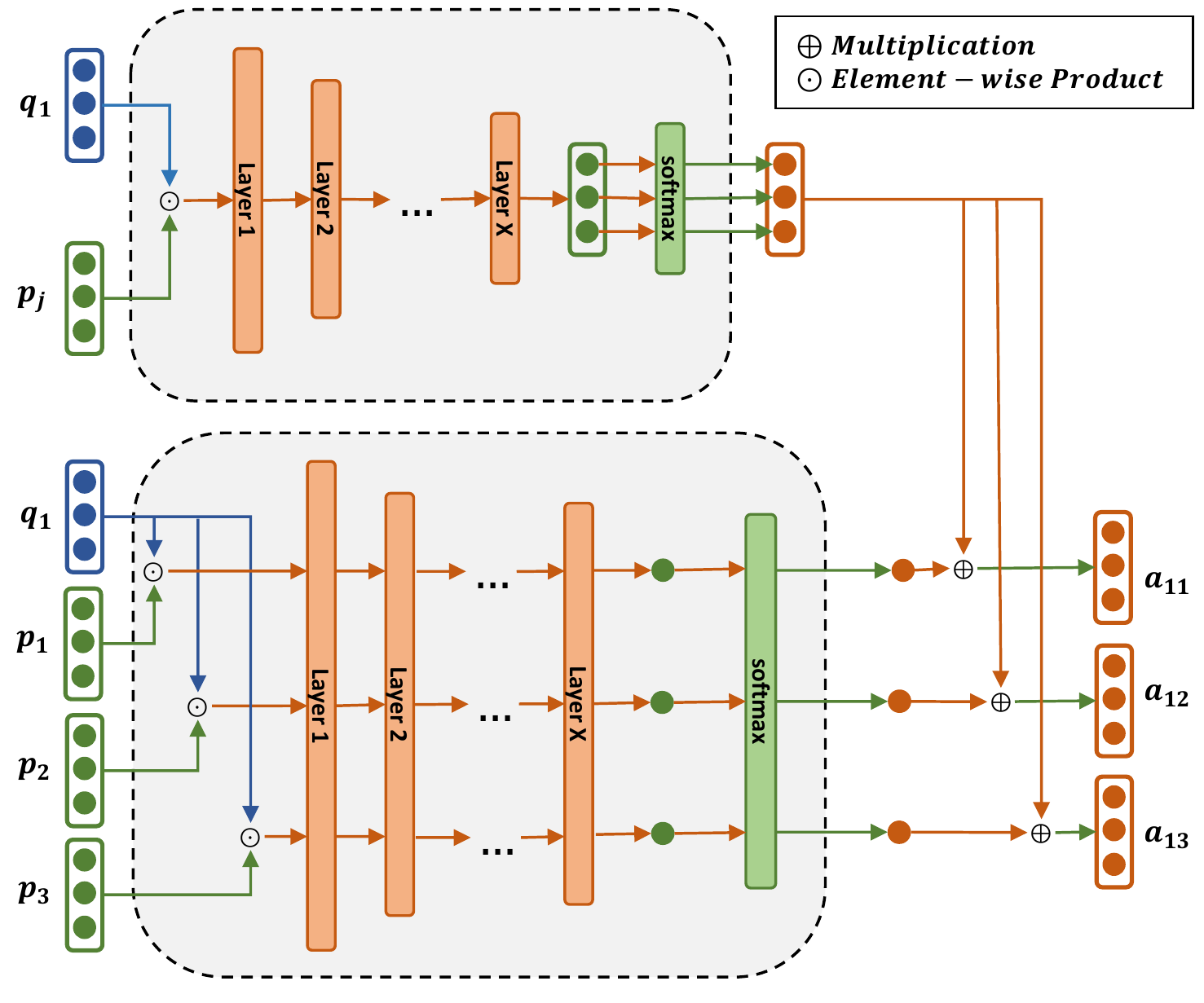}
    \caption{The structure of the item- and feature-level attention network for the first design.}
    \label{fig3}
\end{figure}

\textbf{Design 1}. To integrate both the item-level and feature-level attention in an ICF model, an straightforward method is first to use two attention networks to compute the two types of attention separately, and then combine them together.  Fig.~\ref{fig3} shows the network of this design.  Specifically, one network is to model the different importance of previous items, and the other network is to capture the different contributions of features inside items. For the item-level attention, the attentive weight $b_{ij}$ of a historical item $j$ to a target item $i$ is computed according to the method described in NAIS, and we use the element-wise product method in our implementation\footnote{Note that the concatenation method can be also applied here. We use the element-wise product for the ease of computing both the item-level attention and feature-level attention}. Formally, the $b_{ij}$ is computed as following:
\begin{equation}
   v_{ij} =\mathbf{h}^T ReLU(\mathbf{W}(\mathbf{p}_i\odot\mathbf{q}_j)+\mathbf{b}),
\end{equation}
\begin{equation} \label{eq:tstepsoft}
   b_{ij}=\frac{\exp(v_{ij})}{[\sum_{j'\in\mathcal{R}_u^+\backslash\{i\}}\exp( v_{ij'})]^\beta},.
\end{equation}
Here the parameters $\mathbf{W}$, $\mathbf{b}$, $\mathbf{h}$ and the variant of softmax function (Eq.~\ref{eq:tstepsoft}) are defined as they are in NAIS. For the feature-level attention, the attentive weight vector $\mathbf{\hat{a}_{ij}}$ is computed based on the Eq.~\ref{eq:fla} and ~\ref{eq:softmax}. The final attentive weight of a feature of the item $j$ is weighted by the item attentive weight.
\begin{equation} \label{eq:comb1}
   \mathbf{a_{ij}} = b_{ij}\cdot \mathbf{\hat{a}_{ij}}.
\end{equation}
It is easy to understand the intuition of the equation. If the historical item itself is irrelevant to the target item, the impact of its features on predicting user preference to the target item should also be small. This method is easy to understand but the network structure is complicated.
%This method is easy to understand but it is computationally complex and cumbersome. Since the final result is still an attentive weight vector for each historical item, we introduce a fusion method, which does not need to compute the item- and feature-level attention separately first.

\begin{figure}[t]
    \centering
    \includegraphics[width=0.8\textwidth]{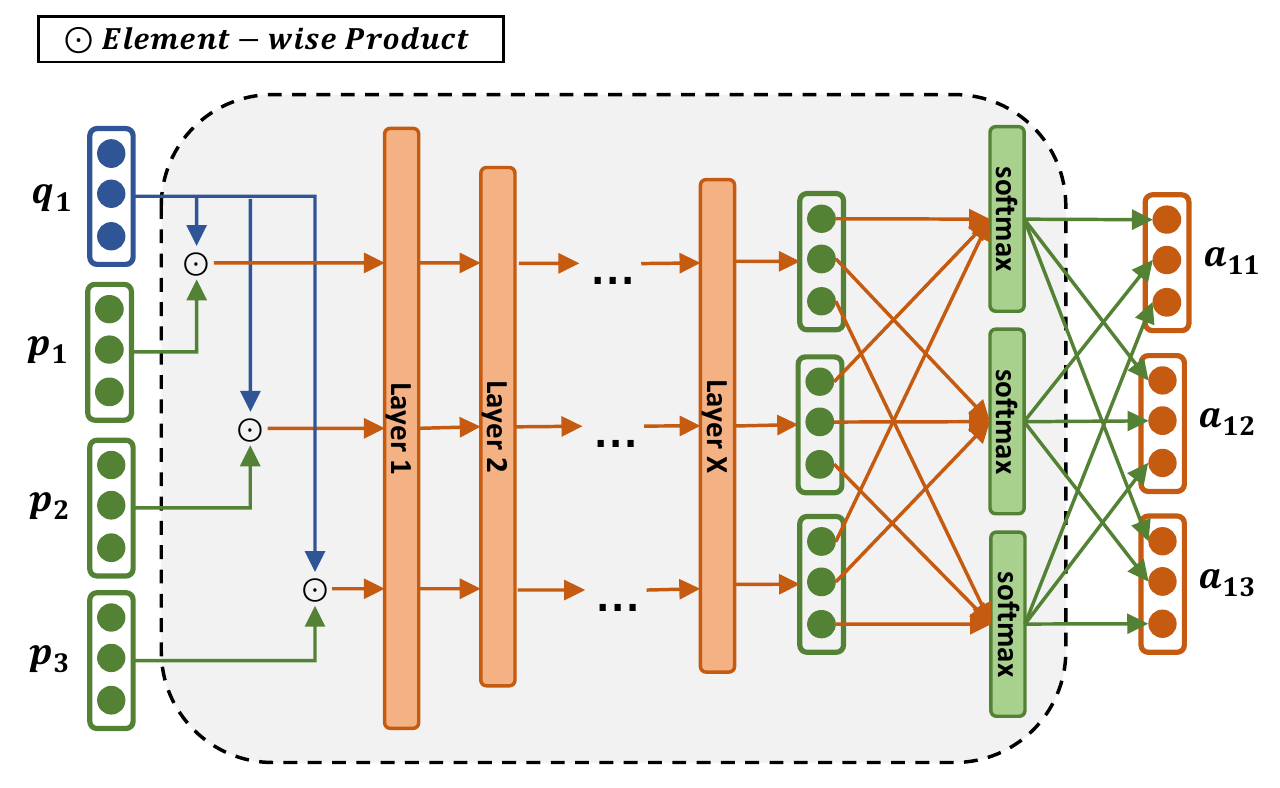}
    \caption{The structure of the item- and feature-level attention network for the second design.}
    \label{fig4}
\end{figure}

\textbf{Design 2.} Since our goal is still to assign an attentive weight vector for each historical item, we attempt to simplify the network structure in the first design and propose a fusion method, whose structure is shown in Fig.~\ref{fig4}. In this method, the attentive weights of a historical item $j$'s features for a target item are computed as:
%     \hat{y}_{ui}=\mathbf{q}_i^\top(\sum_{j\in\mathcal{R}_u^+\backslash\{i\}}\mathbf{a}_{ij}\odot\mathbf{p}_j).
% \end{equation}
\begin{equation}
    \begin{cases}
        \mathbf{\hat{a}_{ij}}=\mathbf{H}^\top ReLU(\mathbf{W}(\mathbf{p}_i\odot\mathbf{q}_j)+\mathbf{b}) \\
        a_{ijk}=softmax'(\mathbf{f}(\hat{a}_{ijk})), ~~~ \text{    } k=1,2,\cdots,d \\
    \end{cases}
\end{equation}
Similar to the calculation of the feature-level attention, we first model the interactions between the historical items and the target items using a nonlinear transformation upon the element-wise product of their embedding vectors. The computation of $\mathbf{\hat{a}_{ij}}$ is the same as in Eq.~\ref{eq:fla} and the notations are defined in the same way. The difference comes from the normalization part. For the feature-level attention inside an item in the ``Design 1", the attentive weight of a feature is normalized over the weights of \emph{all the features of this item}. In this design, the weight of a feature is normalized over the weights of \emph{all the historical items on this particular feature}. In particular, the final attentive weight of the $k$-th feature inside an item $j$ is obtained via a normalization based on the variant of the softmax function~\cite{He2018Nais}:

\begin{equation} \label{eq:comb2}
    softmax'(\mathbf{\hat{a}_{ijk}})=\frac{\exp(a_{ijk})}{[\sum_{j'\in\mathcal{R}_u^+\backslash\{i\}}\exp(\hat{a}_{ij'k})]^\beta}.
\end{equation}
Comparing to the Eq.~\ref{eq:softmax}, we can see that for a feature of a historical item, its importance is evaluated among the same feature of all the historical items in the normalization. In this way, the computation of the feature-level takes the item-level effects into consideration. Notice that it is possible that the attentive weights of all the features of an item are small because this item is not relevant to the target item. Similar in NAIS, the hyper-parameter $\beta$ is to smooth the value of the denominator in softmax. It can help regulate the weights of the item features for users with different numbers of interacted items.

The mechanisms of the above two designs for considering both the item- and feature-level attentions are different. It is theoretically difficult to analyze which one works better in practice. The advantage of the second method is that the network structure is simple and computationally efficient. Besides, it has less parameters and thus is relatively more resistant to overfitting over the first method. We compare the recommendation performance of the two methods in experiments.

\subsection{Prediction}
With the attentive weight vector for each historical item $j \in \mathcal{R}_u$ of a user $u$, the preference to a target item $i$ based on the feature-level attentive method is predicted by:

 \begin{equation} \label{eq:flar}
    \hat{r}_{ui}=\sum_{j\in\mathcal{R}_u^+\backslash\{i\}}\mathbf{p}_i^T(\mathbf{a_{ij}} \odot \mathbf{q}_j).
\end{equation}

From this equation, we can see that our model considers the influence of different features of all the historical items for the preference prediction.

\section{Feature-level Attention Enhanced ICF Models} \label{sec:app}
The proposed feature-level attention model can be easily applied to existing embedding-based ICF models. In this section, we show the applications of our feature-level attentive (FLA) method to two recently proposed ICF models: NAIS~\cite{He2018Nais} and DeepICF~\cite{Xue2019Deep}. For the ease of presentation, we name the two models with the use of our feature-level attention method as \textbf{FLA$_{NAIS}$} and \textbf{FLA$_{DICF}$}, respectively.

%\textbf{FLA$_{FISM}$}. The original FISM treats all the historical items equally for the preference prediction of the target item. The feature-level attention can be directly integrated into the FISM by first using Eq.~\ref{eq:fla} \& Eq.~\ref{eq:softmax} to compute the feature-level attentive weights, and then Eq.~\ref{eq:flar} for preference prediction. Notice that we can also apply the integration of the item- and feature-level attention to FISM. Here we only consider the feature-level attention because of two reasons: (1) only considering the feature-level attention in FLA$_{FISM}$, we can analyze the effects of considering feature-level attention alone in recommendation by comparing its performance to FISM; and (2) the integration of the item- and feature-level attention is actually the application of feature-level attention in NAIS, which is introduced below.

\subsection{FLA$_{NAIS}$: FLA-enhanced NAIS method}
\textbf{NAIS.} We first briefly describe the NAIS method and then introduce the FLA$_{NAIS}$ method. NAIS introduces the attention mechanism~\cite{Chen2017Attentive} to assign different weights to different items. Specifically,  the prediction of NAIS is formulated as:
 \begin{equation}
    \hat{r}_{ui}=\sum_{j\in\mathcal{R}_u^+\backslash\{i\}}a_{ij}\mathbf{p}_i^T\mathbf{q}_j,
\end{equation}
where $a_{ij}$ denotes the attentive weight assigned to the similarity $s_{ij}$, indicating  the contribution of item $j$ to the preference prediction of item $i$. The attentive neural network is used to automatically learn $a_{ij}$ by taking $\mathbf{p}_i$ and $\mathbf{q}_i$ as input. Two different methods have been presented in NAIS to combine $\mathbf{p}_i$ and $\mathbf{q}_i$, i.e., vector concatenation and element-wise product:
\begin{equation} \label{eq:nais}
     \begin{cases}
         f_{concat}(\mathbf{p}_i,\mathbf{q}_j)=\mathbf{h}^T ReLU(\mathbf{W}
         \begin{bmatrix}
             \mathbf{p}_i \\
             \mathbf{q}_j
         \end{bmatrix}
         +\mathbf{b}) \\
         f_{prod}(\mathbf{p}_i,\mathbf{q}_j)=\mathbf{h}^T ReLU(\mathbf{W}(\mathbf{p}_i\odot\mathbf{q}_j)+\mathbf{b}),
    \end{cases}
 \end{equation}
where $\mathbf{W}\in \mathds{R}^{d' \times d}$ and $\mathbf{b}\in \mathds{R}^{d'}$ represent the weight matrix and bias vector of the attention network, respectively. $d'$ denotes the size of the hidden layer. $\mathbf{h}$ is the weight vector of the output layer of the attention network.  $ReLU$~\cite{maas2013rectifier} is used as the activation function.
The $a_{ij}$ is then normalized via a modified softmax function:
 \begin{equation} \label{eq:soft}
    a_{ij}=\frac{\exp(f(\mathbf{p}_i,\mathbf{q}_j))}{[\sum_{j\in\mathcal{R}_u^+\backslash\{i\}}\exp(f(\mathbf{p}_i,\mathbf{q}_j))]^\beta},
 \end{equation}
 where $\beta$ is a hyperparameter to smooth the denominator of the softmax function. The rational lies in the fact that the number of users' interacted items can vary in a wide range. As a result, the standard softmax normalization will overly punish the weights of active users, who have much more interacted items than inactive users.  With a smaller $\beta$, the denominator can be suppressed and thus reduce the punishment on the attention weights of active users~\cite{He2018Nais}. Notice that with the normalization of the modified softmax, the normalization term ($\frac{1}{(|\mathcal{R}u^+|-1)^\alpha}$) is discarded in NAIS.

\textbf{FLA$_{NAIS}$.}
NAIS~\cite{He2018Nais} considers the different contributions of historical items. The application of the feature-level attention to NAIS is the integration of the item- and feature-level. Therefore, the methods described in section~\ref{sec:ifla} is applied to compute the attentive weight vectors for historical items (Eq.~\ref{eq:comb1} or Eq.~\ref{eq:comb2}), and then the preference to the target item is predicted by Eq.~\ref{eq:flar}.

\subsection{FLA$_{DICF}$: FLA-enhanced DeepICF method}

\textbf{DeepICF.}  DeepICF also considers the item-level attention. It first adopts a pairwise interaction layer to model the interaction between each historical item and the target item, and then introduces an attention-based pooling layer to assign different weights to the outputs (of the pairwise interaction layer) from different historical layer. In the next, the output from the previous two layers are fed into deep interaction layers, which consist of a multi-layer perceptron, to model the high-order interaction between items. Finally, a linear regression model is applied to predict the preference. In the next, we introduce each component in details for a clear impression of the DeepICF model. The pairwise interaction layer and the attention-based pooling layer are expressed as:
\begin{equation} \label{eq:eui}
    \mathbf{e_{ui}} = \sum_{j\in\mathcal{R}_u^{+}\backslash\{i\}}a_{ij} (\mathbf{p}_i \odot \mathbf{q}_j),
\end{equation}
where $\odot$ indicates element-wise product. $a_{ij}$ is the item-level attention, which denotes the contribution of item $j$ to the user preference on item $i$.  The attention is computed as:
\begin{equation}
    a_{ij} = softmax'(\mathbf{h^T}ReLU(\mathbf{W}(\mathbf{p}_i \odot \mathbf{q}_j) + \mathbf{b})),
\end{equation}
where  $\mathbf{W}$, $\mathbf{b}$, and $\mathbf{h}$ are defined as in Eq.~\ref{eq:nais} represent the weight matrix and bias vector of the attention network, respectively. $softmax'$ is a modified softmax function as defined in NAIS (see Eq.~\ref{eq:soft}). The deep interaction layers are stacked above the output of the interaction layer to model the higher-order item relations as follows:
\begin{equation}
  \mathbf{e_L} = ReLU(\mathbf{W_L}(ReLU(\mathbf{W_{L-1}} \cdots ReLU(\mathbf{W_1e_{ui}}+\mathbf{b_1}))+\mathbf{b_{L-1}})+\mathbf{b_L}),
\end{equation}
where $\mathbf{W_l}$, $\mathbf{b_l}$, and $\mathbf{e_l}$ denote the weight matrix, bias vector, and output vector of the $l$th hidden layer respectively. $L$ is the total number of network layers. Finally, the prediction is achieved by a linear regression model:
\begin{equation} \label{eq:reg}
   \hat{r}_{ui}=\mathbf{V}^T\mathbf{e_L} + b_{u} + b_i,
\end{equation}
where $\mathbf{V}$ is the weight vector for the prediction; $b_u$ and $b_i$ are the user and item bias as in the standard matrix factorization~\cite{Koren2009Matrix}.

\textbf{FLA$_{DICF}$.}
Similar to NAIS, the methods in section~\ref{sec:ifla} is used to compute the attentive weight vectors for historical item  model. The attentive weight ($a_{ij}$) in Eq.~\ref{eq:eui} is replaced by the obtained weight vector  ($\mathbf{a_{ij}}$), and Eq.~\ref{eq:eui} becomes:
\begin{equation}
    \mathbf{e_{ui}} = \sum_{j\in\mathcal{R}_u^{+}\backslash\{i\}} \mathbf{p}_i \odot (\mathbf{a_{ij}}\odot \mathbf{q}_j)
\end{equation}
We keep the other parts as the same as DeepICF, and thus the preference is still predicted by Eq.~\ref{eq:reg}

\subsection{Optimization}
% To learn the parameters of our model, we leverage the point-wise~\cite{He2017Neural,Li2015Deep} log loss for training.
In this work, we target at the top-$n$ recommendation,  which is a more practical task than rating prediction in real commercial systems~\cite{Rendle2009Bpr}. It aims to  recommend a set of $n$ top-ranked items which match the target user's preferences. Similar to other rank-oriented recommendation work~\cite{He2018Nais,wang2019ngcf,He2020lightgcn}, we adopt the pairwise-based learning method for optimization. As we would like to validate the effectiveness of the proposed feature-level attention, we strictly follow the  implicit feedback setting in the work of NAIS~\cite{He2018Nais} and DeepICF~\cite{Xue2019Deep}, where each user-item interaction has a value of 1 and other non-observed user-item pairs have a 0 value. The recommendation model is also treated as a binary classification task, and the objective function is as follows:
\begin{equation}
    L=-\frac{1}{|\mathcal{R}^+|+|\mathcal{R}^-|}\left(\sum_{(u,i)\in\mathcal{R}^+}\log\sigma(\hat{r}_{ui})+\sum_{(u,i)\in\mathcal{R}^-}\log(1-\sigma(\hat{r}_{ui}))\right)+\lambda\Vert\Theta\Vert^2,
\end{equation}
where $\mathcal{R}^+$ denotes the positive instances set and $\mathcal{R}^-$ denotes the negative one where each user-item instance is sampled from the non-interacted pairs;  $\sigma$ is a sigmoid function, which can convert the predicted score $\hat{r}_{ui}$ of user $u$ and item $i$ into a probability representation, constraining the result to (0,1);  $\lambda$ is the parameter to control the effect of $\ell_2$ regularization, which is used to prevent overfitting; and $\Theta$ represents all the trainable parameters including $\mathbf{p}_i$, $\mathbf{q}_j$, $\mathbf{H}$, $\mathbf{W}$ and $\mathbf{b}$. In addition, FLA$_{DICF}$ has a multi-layer perception behind the attention network to simulate high-level interactions of users and $\Theta$ also contains their weight parameters.

\textbf{Model training}.
We adopted Adagrad~\cite{Duchi2011Adaptive} to optimize the prediction model and update the model parameters. Because the objective function is non-convex,  the loss function might be trapped in a local minimum, resulting in sub-optimal performance. Previous work has demonstrated that \emph{pre-training} is particularly useful in practice for accelerating the training process and achieving better performance~\cite{He2017Neural,He2018Adversarial}.  We will report the results with and without the pre-training in experiments (see section~\ref{sec:pre-train}).

\section{Experiments} \label{sec5}
We conducted extensive experiments on six publicly accessible datasets to evaluate the effectiveness of the proposed method. In particular, we mainly answer the following research questions.

\begin{itemize} [align=left,style=nextline,leftmargin=*,labelsep=\parindent,font=\normalfont]
\item \textbf{RQ1:} Which design is better to integrate the item-level and feature-level attention, \emph{Design 1} or \emph{Design 2}?

\item \textbf{RQ2:} Are our proposed feature-level attention methods useful for providing more accurate recommendations? How do our feature-level attention enhanced methods perform with comparison to the state-of-the-art item-based CF methods?

\item \textbf{RQ3:} How do the hyper-parameters, i.e.,  the embedding size $d$ and $\beta$, affect the performance of the feature-level attention enhanced methods?

\item \textbf{RQ4:} Is the pre-training strategy useful for our feature-level attention enhanced methods?

\end{itemize}

In what follows, we first presented the experimental settings, and then answered the above questions sequentially based on experimental results.

\subsection{Experimental Setup}
\textbf{Datasets.} We evaluated on the following publicly accessible datasets: MovieLens\footnote{https://grouplens.org/datasets/movielens/1m/.}, Delicious\footnote{https://grouplens.org/datasets/hetrec-2011/.}, and Amazon review datasets\footnote{https://jmcauley.ucsd.edu/data/amazon/.}.

\begin{itemize}[align=left,style=nextline,leftmargin=*,labelsep=\parindent,font=\normalfont]
\item \textbf{MovieLens (ML-1m).}  This movie rating dataset has been widely adopted to evaluate recommendation algorithms. We used the version containing one million ratings, where each user has at least 20 ratings.

\item \textbf{Delicious.} This dataset contains social networking, bookmarking, and tagging information  from a set of 2K users from Delicious social bookmarking system\footnote{http://www.delicious.com. }.

\item \textbf{Amazon.} This dataset contains user interactions on items as well as item metadata from Amazon~\cite{mcauley2013hidden}. In our experiments, we only used the interaction information. For each observed user-item interaction, we treated it as a positive instance; otherwise, it is negative. Four product categories from this dataset are used in evaluation: Music, Beauty, CDs, and Movies. We downloaded the 5-core  version,  which means that users and items have at least 5 interactions.  To make our datasets more diverse in density, we kept the 5-core version for the Music and Beauty datasets and  further processed the CDs and Movies datasets to ensure that each user and item have at least 10 interactions.
\end{itemize}

 The basic statistics of the used datasets are shown in Table~\ref{tab:data}. As we can see, the selected datasets are of different sizes and sparsity levels. For example, ML-1m is the most denser dataset, and the Amazon datasets are quite sparse. This can help us evaluate the performance of adopted methods for item recommendation under different scenarios in experiments.
\begin{table}[t]
    \caption{Basic statistics of the experimental datasets.}
    \centering
    \begin{tabular}{|c|c|c|c|c|}
        \hline
        \textbf{Dataset} & \textbf{\#Users} & \textbf{\#Items} & \textbf{\#Ratings} & \textbf{Sparsity}\\
        \hline\hline
        ML-1m & 6,040 & 3,416 & 999,611 & 95.16\%\\
        \hline
        Delicious & 1,055 & 1,055 & 11,784 & 98.94\%\\
        \hline
        Music & 5,541 & 3,568 & 64,706 & 99.67\%\\
        \hline
        Beauty & 22,363 & 12,101 & 198,502 & 99.93\%\\
        \hline
        CDs & 15,592 & 16,184 & 445,412 & 99.82\%\\
        \hline
        Movies & 33,326 & 21,901 & 958,986 & 99.87\%\\
        \hline
    \end{tabular}
    \label{tab:data}
\end{table}

\textbf{Evaluation Protocols.} We focused on the top-$n$ recommendation task, aiming to recommend a set of $n$ top-ranked items that will be appealing to the target user. For each dataset, we randomly splitted it into training, validation, and testing set with the ratio 70:10:20 for each user. The observed user-item interactions were treated as positive instances. The performance was evaluated by the widely used metrics - \emph{Hit Ratio} (HR)~\cite{deshpande2004hr} and \emph{Normalized Discounted Cumulative Gain} (NDCG)~\cite{deshpande2004hr}. For each metric, the performance of recommendation methods is often judged by the top $n$ results.  Particularly, $HR@n$ is a recall-based metric, measuring whether the test item is in the top-$n$ positions of the recommendation list. $NDCG@n$ emphasizes the quality of ranking, which assigns higher score to the top-ranked items by taking the position of correctly recommended into considerations.  The reported results are the average values across all the tested users based on the top 10 results (i.e., $n=10$).

\textbf{Compared Baselines.}
As the main contribution of this work is to advocate the importance of considering the feature-level attention in recommendation, especially for the item-based CF recommendation methods. Therefore,  we mainly compared our feature-level attention enhanced methods with the state-of-the-art ICF models in the empirical studies. Specifically, we compared our methods with the following baselines.

    \begin{itemize}[align=left,style=nextline,leftmargin=*,labelsep=\parindent,font=\normalfont]
        \item \textbf{Random} recommends items randomly to users.

        \item \textbf{Pop} represents the popularity-based method, which recommends items according to their popularity, namely, the number of interactions in the dataset.

        \item \textbf{ItemKNN} is the standard item-based collaborative filtering method~\cite{deshpande2004hr}.

        \item \textbf{BMF} is the standard matrix factorization method with the consideration of bias terms~\cite{Koren2009Matrix}. It is originally designed for rating prediction. We adapt it for the top-$n$ recommendation task by ranking items based on the predicted ratings.

        \item \textbf{BPR}~\cite{Rendle2009Bpr} is a popular pair-wise learning method, which employs a Bayesian Personalized Ranking loss to optimize the matrix factorization model. This is a basic baseline with competitive performance on the top-$n$ recommendation task and has been widely used in empirical studies to evaluate the newly developed method.

        \item \textbf{MLP}~\cite{He2017Neural} uses a multi-layer perceptron above user and item embeddings to replace the inner product for recommendation. Following the setting in~\cite{He2018Nais}, we also use a three-layer MLP and optimize the point-wise log loss in experiments.

        %\item \textbf{SLIM}~\cite{Ning2011Slim} is the earliest learning-based item-based CF model. It learns an item-item similarity matrix to reconstruct the user-item interaction function.

        \item \textbf{FISM}~\cite{Kabbur2013Fism} is a pioneering learning-based ICF model by directly learning item embeddings. In experiments, we carefully tuned $\alpha$ (see Eq.(3) in~\cite{He2018Nais}) from 0 to 1 with a step size of 0.1 and reported the best result for each experimental dataset.

        \item \textbf{NAIS}~\cite{He2018Nais} is a state-of-the-art item-based CF model. It considers the different effects of historical items to the target item and applies the attention mechanism to model the item-level attention in prediction.

        \item \textbf{DeepICF}~\cite{Xue2019Deep} is a recently proposed deep ICF method which can capture the high-order interactions between items. By stacking multiple perceptron layers above the interactions between items, it adopts a linear regression for the final prediction.
    \end{itemize}

Random, Pop, ItemKNN are basic recommendation methods without learning user and item embeddings. BMF learns user and item embeddings by reconstructed the user-item rating matrix. The objective  function is to minimize the re-constructed errors, and thus it is not rank-oriented.  BRP is a traditional and competitive CF model based on matrix factorization for the top-$n$ recommendation task. MLP is a state-of-the-art CF method based on the neural network proposed in recent years. Both methods are widely used as baselines in many studies, and they are used as the basic references to show the performance of other methods. FISM is a representative learning-based ICF method. NAIS and DeepICF are the main baselines to compared with  FLA$_{NAIS}$ and FLA$_{DICF}$ to study the effects of our feature-level attention approach in recommendation.

\textbf{Parameter Settings.} For fair comparisons, for the methods using pair-wise learning strategy, we paired each positive instance in the training set with four randomly sampled negative instances to train all methods. Four embedding sizes ($d \in \{8, 16, 32, 64\}$)  are tested in experiments. The learning rate is searched in [0.01, 0.001, 0.0001, 0.00001]. The smoothing parameter $\beta$ is tuned in the range of [0.1, 0.9] with a step size of 0.2 for NAIS, DeepICF, and our methods.  The best results of each method on the test datasets are reported below. Without particular specifying the value of $\beta$, the reported results are obtained by setting $\beta=0.7$ for all the models. It is also worth mentioning that we used the learned user/items' embeddings by FISM as model initialization for NAIS, FLA$_{NAIS}$, DeepICF, and FLA$_{DICF}$. The benefits of pre-training for NAIS and DeepICF have been demonstrated in~\cite{He2018Nais} and ~\cite{Xue2019Deep}, respectively. We also study its effects to FLA$_{NAIS}$ and FLA$_{DICF}$ in Section~\ref{sec:pre-train}.

\subsection{Performance Comparisons of Different Designs (RQ1)}
In this section, we reported the performance of different designs for combining item-level and feature-level attentions in ICF models, namely, \emph{Design 1} and \emph{Design 2} as described in Section~\ref{sec4}. Table~\ref{tab2} shows the performance (in terms of HR@10 and NDCG@10) of applying the two designs to NAIS and DeepICF (i.e., FLA$_{NAIS}$ and FLA$_{DICF}$) on the six evaluation datasets. From the results, we can see that for both NAIS and DeepICF models, the \emph{Design 2} can obtain consistently and slightly better results than the \emph{Design 1}. The better performance of \emph{Design 2} is largely attributed to its simple design with less parameters, making the model easier to be trained and more resistant to overfitting.  Because of the better performance of \emph{Design 2} in our experiments, in the following sections, all the reported results of FLA$_{NAIS}$ and FLA$_{DICF}$ are based on \emph{Design 2}.

\begin{table}[t]
    \caption{Performance comparisons between the two designs of our feature-level attention framework. The results are obtained with embedding size 16. Noticed that the values are reported by percentage with `\%’ omitted. }
     \centering
    \resizebox{.99\textwidth}{!}{
        \begin{tabular}{|c|c|c|c|c|c|c|c|c|c|c|c|c|c|c|c|c|c|c|}
            \hline
            \multicolumn{2}{|c|}{\multirow{2}*{\textbf{Methods}}} &
            \multicolumn{2}{c|}{\textbf{ML-1m}} &
            \multicolumn{2}{c|}{\textbf{Delicious}} & \multicolumn{2}{c|}{\textbf{Music}} & \multicolumn{2}{c|}{\textbf{Beauty}} & \multicolumn{2}{c|}{\textbf{CDs}} & \multicolumn{2}{c|}{\textbf{Movies}} \\
            \cline{3-14}
            \multicolumn{2}{|c|}{}  & \textbf{HR} & \textbf{NDCG} & \textbf{HR} & \textbf{NDCG} & \textbf{HR} & \textbf{NDCG} & \textbf{HR} & \textbf{NDCG}& \textbf{HR} & \textbf{NDCG} & \textbf{HR} & \textbf{NDCG}\\
            \hline
            % NAIS & 40.21 & \textbf{21.68} & 68.76 & 43.96 & 54.58 & 33.77 & 54.02 & 34.75 & 45.03 & 27.95 & 50.02 & 30.71\\
            \multirow{2}*{\textbf{FLA$_{NAIS}$}} &Design$_1$ &\textbf{90.46} &35.54 & 74.60 & 38.77 &22.97 &8.28 &9.02 &3.48 &23.68 &5.61 &16.04 &3.52 \\
            &Design$_2$ &90.36 &\textbf{36.18} & \textbf{76.11} & \textbf{38.84} &\textbf{23.25} &\textbf{8.55} &\textbf{9.13} &\textbf{3.52} &\textbf{24.04} &\textbf{5.81} &\textbf{16.30} &\textbf{3.58} \\
            \hline
            % FAMR-NAIS & \textbf{40.45} & 21.55 & \textbf{69.32} & \textbf{44.31} & \textbf{55.17} & \textbf{34.25} & \textbf{54.69} & \textbf{34.98} & \textbf{45.69} & \textbf{28.43} & \textbf{50.46} & \textbf{30.82}\\

            \multirow{2}*{\textbf{FLA$_{DICF}$}} &Design$_1$ &\textbf{90.21} &35.52 & 73.65 & 37.58 &21.17 &8.04 &7.74 &2.89 &23.44 &5.52 &14.48 &3.35 \\
            &Design$_2$ &90.11 &\textbf{35.73} & \textbf{73.93} & \textbf{38.41} & \textbf{22.22} & \textbf{8.08} & \textbf{8.17} & \textbf{3.03} &\textbf{23.47} & \textbf{5.65} & \textbf{15.59} & \textbf{3.43} \\
            \hline
        \end{tabular}
    }
    \label{tab2}
\end{table}

\begin{table}[t]
    \caption{Performance of HR@10 and NDCG@10 of compared approaches at embedding size 16. Noticed that the values are reported by percentage with `\%’ omitted. The symbols (``*") and (``$\Delta$") after a numeric value denotes significant differences based on a two-tailed paired t-test of our FLA-enhanced ICF methods and their counterparts with $p < 0.05$ and $p < 0.01$, respectively.  }
     \centering
    \resizebox{.99\textwidth}{!}{
        \begin{tabular}{|c|c|c|c|c|c|c|c|c|c|c|c|c|c|c|c|c|}
            \hline
            \multirow{2}*{\textbf{Methods}} &
            \multicolumn{2}{c|}{\textbf{ML-1m}} &
            \multicolumn{2}{c|}{\textbf{Delicious}} & \multicolumn{2}{c|}{\textbf{Music}} & \multicolumn{2}{c|}{\textbf{Beauty}} & \multicolumn{2}{c|}{\textbf{CDs}} & \multicolumn{2}{c|}{\textbf{Movies}} \\
            \cline{2-13}
            ~ & \textbf{HR} & \textbf{NDCG} & \textbf{HR} & \textbf{NDCG} & \textbf{HR} & \textbf{NDCG} & \textbf{HR} & \textbf{NDCG}& \textbf{HR} & \textbf{NDCG} & \textbf{HR} & \textbf{NDCG}\\
            \hline\hline
            Random &10.18 &1.20 &1.33 &0.39 &0.58 &0.13 &0.16 &0.06 &0.38 &0.05 &0.28 &0.04  \\ \hline
            Pop &55.70 &9.93 &8.16 &2.20 &3.56 &1.24 &1.91 &0.74 &3.41 &0.64 &3.45 &0.68 \\ \hline
            ItemKNN &75.63 &26.08 &11.71 &3.71 &4.38 &1.43 &2.16 &0.96 &3.66 &0.76 &3.77 &0.74 \\ \hline \hline
            BMF &87.73 &37.59 &16.30 &7.50 &5.99 &2.10 &2.57 &1.06 &4.66 &0.91 &4.51 &0.91  \\ \hline
            BPR &70.40 &22.59 & 68.69 & 40.79 &19.00 &7.70 &6.65 &3.06 &11.13 &3.78 &10.75 &2.51 \\
            \hline
            MLP &85.36 &31.32 & 70.71 & 41.61 &11.10 &3.67 &6.25 &2.38 &9.53 &3.04 &6.02 &1.84 \\
            \hline
            FISM &89.37 &33.50 & 72.70 & 37.04 &21.08 &7.59 &7.83 &2.93 &21.25 &4.94 &14.69 &3.13 \\
            % \hline
            % FAMR-FISM & 34.40 & 18.28 & 59.88 & 36.08 & 49.62 & 29.27 & 50.20 & 30.37 & 35.85 & 20.83 & 49.22 & 29.52\\
            \hline\hline
            % NAIS & 40.21 & \textbf{21.68} & 68.76 & 43.96 & 54.58 & 33.77 & 54.02 & 34.75 & 45.03 & 27.95 & 50.02 & 30.71\\
            NAIS &89.77 &34.03 & 75.26 & 37.88 &22.31 &8.27 &9.02 &3.47 &23.44 &5.52 &15.81 &3.42 \\
            \hline
            % FAMR-NAIS & \textbf{40.45} & 21.55 & \textbf{69.32} & \textbf{44.31} & \textbf{55.17} & \textbf{34.25} & \textbf{54.69} & \textbf{34.98} & \textbf{45.69} & \textbf{28.43} & \textbf{50.46} & \textbf{30.82}\\
            FLA$_{NAIS}$ &90.36$^\Delta$ &36.18$^\Delta$ & 76.11$^\Delta$ & 38.84* &23.25* &8.55* &9.13* &3.52 &24.04* &5.81* &16.30$^\Delta$ &3.58$^\Delta$ \\
            \hline\hline
            DeepICF &88.25 &33.78 & 73.74 & 37.56 &21.04 &7.84 &7.85 &2.97 &22.73 &5.39 &14.87 &3.22 \\
            \hline
            FLA$_{DICF}$ &90.11$^\Delta$ &35.73$^\Delta$ & 73.93* & 38.41$^\Delta$ &22.22$^\Delta$ &8.08* &8.17* &3.03* &23.47$^\Delta$ &5.65$^\Delta$ &15.59* &3.43* \\
            \hline
        \end{tabular}
    }
    \label{tab3}
\end{table}
\subsection{Model Comparison (RQ2)}
In this section, we compared the performance of our feature-level attention (FLA) enhanced ICF models with all the adopted competitors. The results of all compared methods over all the test datasets are reported in Table~\ref{tab3} in terms of HR@10 and NDCG@10.  For a fair comparison,  the reported results are based on the same embedding size $d=16$ for all the methods. Because our main goal is to validate the effectiveness of the proposed feature-level attention framework, we performed a statistical significance test with a two-tailed paired t-test between our FLA-enhanced ICF methods and their counterparts (i.e., FLA$_{NAIS}$ v.s. NAIS and  FLA$_{DICF}$ v.s. DeepICF). The symbols (``*") and (``$\Delta$") denote the improvements are significant with $p < 0.05$ and $p < 0.01$, respectively.

First, we would like to validate the effects of feature-level attentions on enhancing the performance of ICF models by comparing FLA$_{NAIS}$ and FLA$_{DICF}$ to NAIS and DeepICF, respectively. From the table, we can  observe that with the consideration of feature-level attentions, the performance of NAIS and DeepICF can achieve better  performance in most cases across the six datasets, which are of different scales and sparsity levels. Note that both NAIS and DeepICF already consider the different importance of items to the target item (i.e., item-level attention), the better performance of FLA$_{NAIS}$ and FLA$_{DICF}$ demonstrates that differentiating the contributions of different features (i.e., feature-level attention) can further improve the performance. The results validate our main assumption that users attend to different features of varied items and the incorporation of feature-level attention into ICF models is beneficial.

NAIS is a direct extension from FISM by considering the item-level attention, we can see that a large improvement of NAIS over FISM. With the additionally considering feature-level attention, FLA$_{NAIS}$ can further improve the performance. Note that NAIS itself is already a very competitive ICF model (with a large improvement over the FISM), and it becomes harder to gain improvement. In addition, our FLA method attempts to capture the feature-level preference of users on items, which needs more interactions or side information to model such a fine-grained level preference on items. By further analyzing the absolute improvements of the FLA-enhanced methods over their counterparts, we can observe a trend that larger improvement can be obtained on denser datasets. Specifically, the absolute improvements in terms of NDCG@10 for FLA$_{NAIS}$ over NAIS are 1.51\%, 0.29\%, and -0.03\% for the the ML-1m, CDs, and Beauty datasets, respectively; and those of FLA$_{DICF}$ over DeepICF are 0.95\%, 0.26\%, and 0.06\% on the three datasets, respectively. The ML-1m is the most denser dataset and Beauty is the most sparse one (see Table 1).   In our experiments, because only the interaction information is exploited and the interactions are often sparse for most users, it is difficult to model the fine-grained preference well, resulting in relatively small improvements. Despite the limited information of the training data, we can still observe a consistent performance improvement by a small  modification of the ICF models - replacing item-level attention (a scalar weight) with our proposed attention network (a weight vector), which is encouraging. Leveraging side information to help capture the feature-level preference of user might be help and we leave it as a future study.

Furthermore, we compared the performance of all the adopted methods. There are some interesting findings: 1) Although BMF is designed for rating prediction, it  still largely outperforms the heuristic-based approaches Pop and ItemKNN, which only perform better than random recommendation. This demonstrates the advantages of learning-based methods.   2) BPR is very competitive when the datasets are relatively denser, such as ML-1m and Delicious. MLP enjoys the advantages of  modeling the non-linear interactions between users and items, but it often requires more training data to model user preference. Therefore, for datasets with richer user interactions, it performances better than BPR, like on the  ML-1m and Delicious datasets. But its performance drops sharply when the data become sparser, resulting in inferior performance on the other four datasets in our experiments.   3) As a learning-based ICF model, FISM achieves better performance  over the above heuristic-based or user-based CF methods.  NAIS consistently outperforms FISM, demonstrating the importance of differentiating the different contributions of items. 4) As discussed above, our FLA-enhanced models achieves the best performance. Note that the performance of the FLA-enhanced models depends on performance of the backbone models. Although FLA$_{DICF}$ obtains much better results than DeepICF, it is still inferior to NAIS in some cases. Overall, the best performance is obtained by FLA$_{NAIS}$, which further enhances the performance of NAIS with the consideration of feature-level attention.

\begin{figure}[t]
    \centering
    \includegraphics[width=\textwidth]{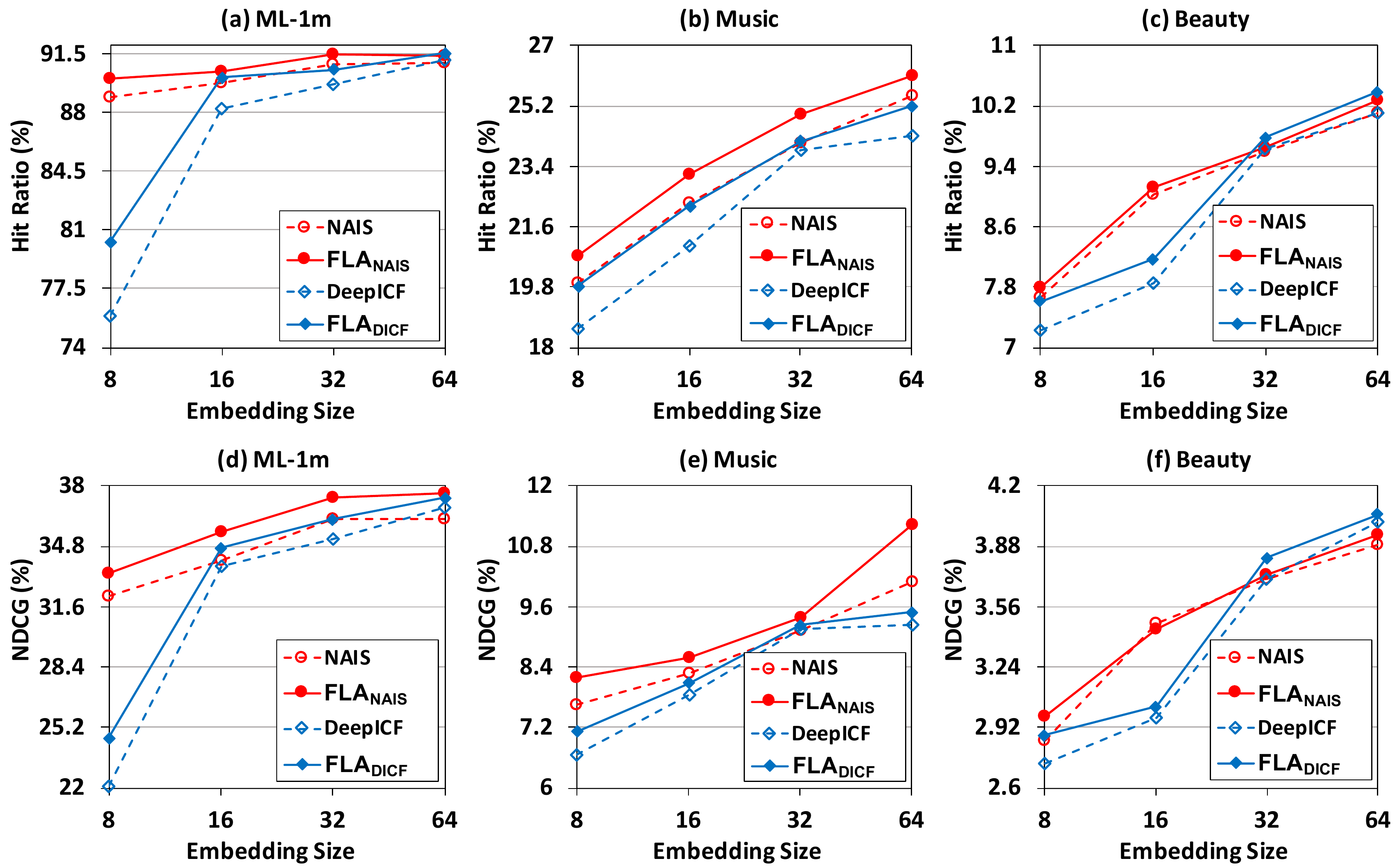}
    \caption{Performance of HR@10 and NDCG@10\emph{ w.r.t.} the number of embedding sizes on three datasets.}
    \label{figrq3_1}
\end{figure}

\subsection{Hyper-parameter Analysis (RQ3)}
In this section, we analyzed the influence of two hyper-parameters, i.e., embedding size $d$ and smoothing exponent $\beta$, on the performance of our feature-level attention enhanced ICF models.

\textbf{Effect of embedding size.}
For analyzing the effect of the embedding size for the performance improvement of the feature-level attention module, we test FLA$_{NAIS}$ and FLA$_{DICF}$ with their counterparts with respect to different embedding sizes. The results on three representative datasets are shown in Figure~\ref{figrq3_1}. The three datasets are selected according to the differences of scales and sparsity levels. Firstly, we can have a clearly observation which is consistent with many previous studies: the performance (in terms of accuracy) of all models is increasing with a larger embedding size, which is attributed to the increasing representation capability of the larger embedding size. Note that when the embedding size continue increasing, there is a risk of overfitting, which has not been observed in this study because the largest embedding size in our experiments is 64. A more interesting observation is that our FLA-enhanced models obtain larger the performance gain with a smaller embedding size.  The underlying reason might be that when the embedding size, i.e., number of item features, is smaller, it is relatively easier for the attention network to learn good feature-level attention weights (because of less features). Therefore, our models can benefits more from the feature-level effects on user preference modeling, leading to better performance.

\begin{figure}[t]
    \centering
    \includegraphics[width=\textwidth]{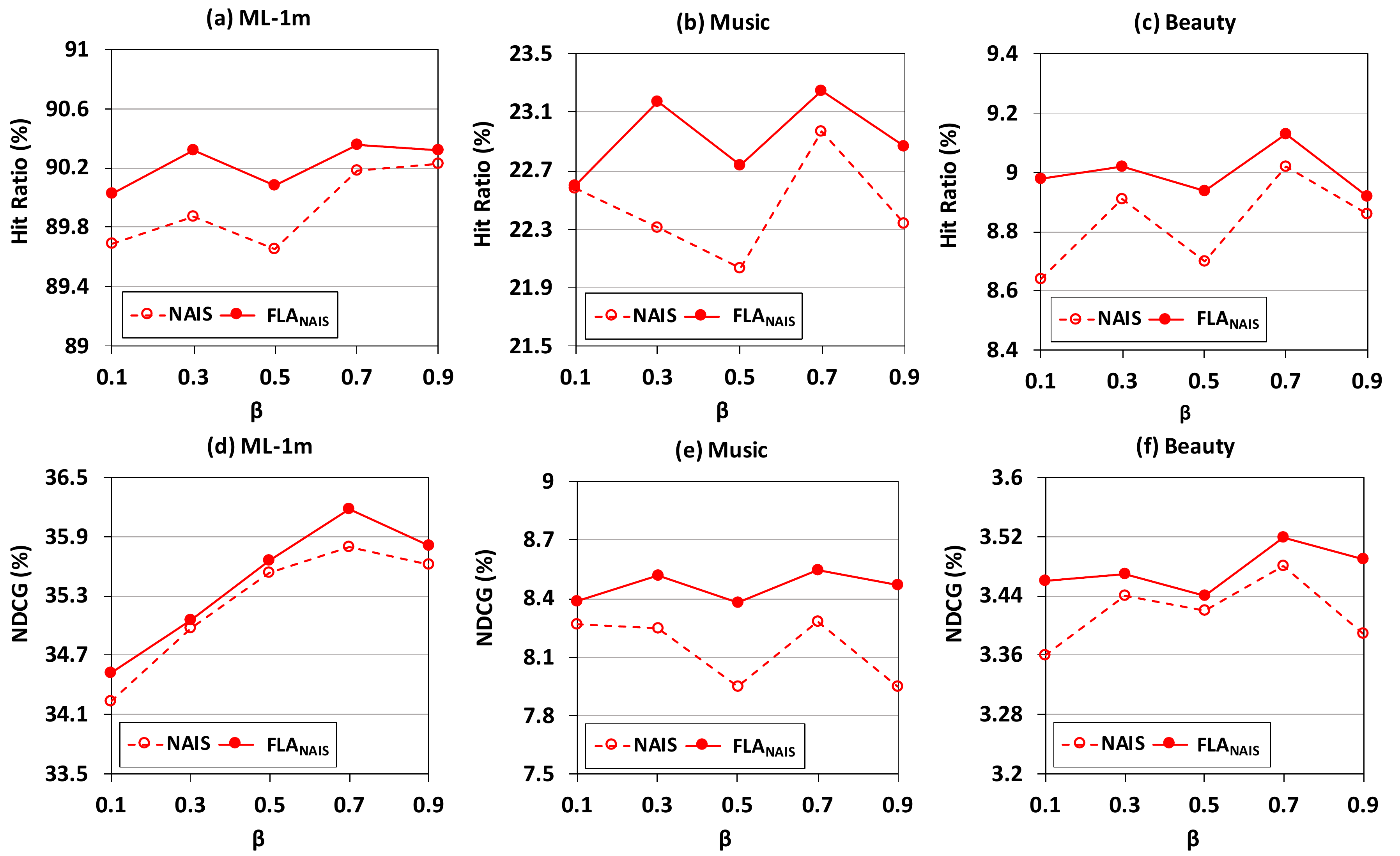}
    \caption{Performance of FAMR-NAIS on different smoothing exponent $\beta$.}
    \label{figrq3_2}
\end{figure}

\textbf{Effect of the smoothing exponent.}
Because of the different numbers of history items for users, using the standard softmax normalization can excessively penalize the weights of active users with a long history. We use the performance of NAIS and FLA$_{NAIS}$ to demonstrate the effects of the smoothing factor $\beta$. We omit results of DeepICF and FLA$_{DICF}$, as they adopted the same smoothing strategy and similar results are observed.  Figure~\ref{figrq3_2} shows the performance of NAIS and FLA$_{NAIS}$ with different $\beta$. We can see FLA$_{NAIS}$ consistently outperforms NAIS; and the general trends of the performance change for both methods are similar, indicating the smoothing effects are the same to the two methods. The optimal value of $\beta$ depends on the target datasets. It seems 0.7 is a good choice across all the datasets. Note that when  $\beta=1$, it means that a standard attention method is used to normalize the attention weights. As pointed out in ~\cite{He2018Nais}, a standard stetting does not work well because of the large variance of the length of user histories. We can observe a dramatic performance degradation when $\beta=0.9$, indicating it already becomes insufficient to reduce the punishment on the attention weights of active users. This also demonstrates the importance of smoothing the denominator in the softmax function for attention weight computation on user behavior data.
% FAMR-h uses the standard softmax function when distributing attention between features, that is, $\beta=1$, and changing the beta has no effect on it. The representation in the figure is a straight line. In addition, FAMR-h only considers the influence of features and does not consider the influence of user history items. As a result, the performance is much lower than the other three methods.
\begin{table}[t]
    \caption{Performance of FLA$_{NAIS}$ and FLA$_{DICF}$ with (/w) and without (/o) pre-training at embedding size 16.}\smallskip
    \centering
    \resizebox{.99\textwidth}{!}{
        \begin{tabular}{|c|c|c|c|c|c|c|c|c|c|c|c|c|c|c|c|c|}
            \hline
            \multirow{2}*{\textbf{Methods}} & \multicolumn{2}{c|}{\textbf{ML-1m}} & \multicolumn{2}{c|}{\textbf{Delicious}} & \multicolumn{2}{c|}{\textbf{Music}} & \multicolumn{2}{c|}{\textbf{Beauty}} & \multicolumn{2}{c|}{\textbf{CDs}} & \multicolumn{2}{c|}{\textbf{Movies}}\\
            \cline{2-13}
            ~ & \textbf{HR} & \textbf{NDCG} & \textbf{HR} & \textbf{NDCG} & \textbf{HR} & \textbf{NDCG} & \textbf{HR} & \textbf{NDCG}& \textbf{HR} & \textbf{NDCG} & \textbf{HR} & \textbf{NDCG}\\
            \hline\hline
            FLA$_{NAIS}$/o & 88.03 & 33.31 & 70.71 & 36.47 & 17.25 & 6.14 & 7.30 & 3.12 & 15.87 & 3.14 & 10.92 & 2.25\\
            \hline
            FLA$_{NAIS}$/w & \textbf{90.36} & \textbf{36.18} & \textbf{76.11} & \textbf{38.84} & \textbf{23.25} & \textbf{8.55} & \textbf{9.13} & \textbf{3.52} & \textbf{24.04} & \textbf{5.81} & \textbf{16.30} & \textbf{3.58}\\
            \hline\hline
            FLA$_{DICF}$/o & 82.42 & 27.58 & 67.87 & 35.30 & 17.89 & 6.13 & 7.42 & 2.81 & 15.68 & 3.49 & 10.13 & 2.06\\
            \hline
            FLA$_{DICF}$/w & \textbf{90.11} & \textbf{35.73} & \textbf{73.93} & \textbf{38.41} & \textbf{22.22} & \textbf{8.08} & \textbf{8.17} & \textbf{3.03} & \textbf{23.47} & \textbf{5.65} & \textbf{15.59} & \textbf{3.43}\\
            \hline
        \end{tabular}
    }
    \label{tab4}
\end{table}
\begin{figure}[t]
    \centering
    \includegraphics[width=\textwidth]{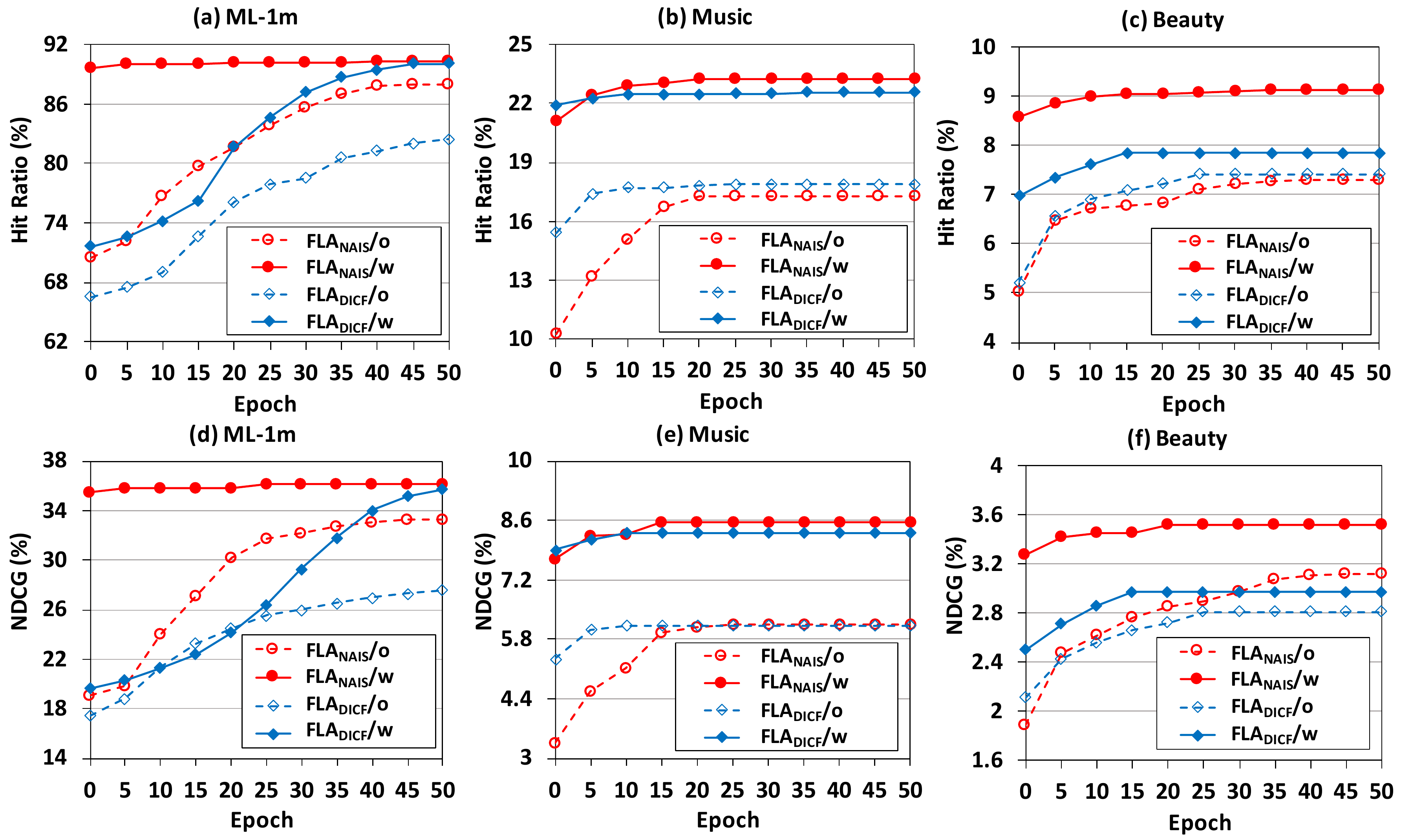}
    \caption{Performance of FLA$_{NAIS}$ and FLA$_{DICF}$ with (/w) and without (/o) pre-training at embedding size 16 at each epoch.}
    \label{figrq4}
\end{figure}

\subsection{Effect of Pre-training (RQ4)} \label{sec:pre-train}
As pre-training has been widely used for model training and demonstrated good performance, we also employ this technique in our experiments. To demonstrate the effects of pre-training, we compare the feature-level attention enhanced models with  (denoted by FLA$_{NAIS}$/w and FLA$_{DICF}$/w) and without pre-training (denoted by FLA$_{NAIS}/o$ and FLA$_{DICF}/o$). In our implementation, we used the learned user/items' embeddings by FISM as model initialization for both FLA$_{NAIS}$ and FLA$_{DICF}$. For FLA$_{NAIS}/o$ and FLA$_{DICF}/o$, the hyper-parameters have been separately tuned. Note that we can also use the learned embeddings by NAIS and DeepICF as model initialization for FLA$_{NAIS}$ and FLA$_{DICF}$. Because NAIS and DeepICF themselves also need pre-training for faster convergence and better performance~\cite{He2018Nais,Xue2019Deep}, it is cumbersome to use their learned embedding in practice. Therefore, we used the embedding learned in FISM as pre-training results for simplicity and consistency. The comparison results with and without pre-training are shown in Table~\ref{tab4}.
It can be seen that with the pre-training, the performance of both methods have been significantly improved. By initializing the model randomly, it is easier to be trapped in local minimums, which hurts the performance of the model. Beyond performance improvements, pre-training can also accelerate the convergence speed. Figure~\ref{figrq4} shows the convergence rate of FLA$_{NAIS}$ with (FAMR/w) and without (FAMR/o) pre-training. We find that in the three datasets \emph{ML-1m}, \emph{Music}, and \emph{Beauty}, there is a faster convergence speed with pre-training than without pre-training.
\subsection{Visualization}
\begin{figure}[t]
    \centering
    \includegraphics[width=0.8\textwidth]{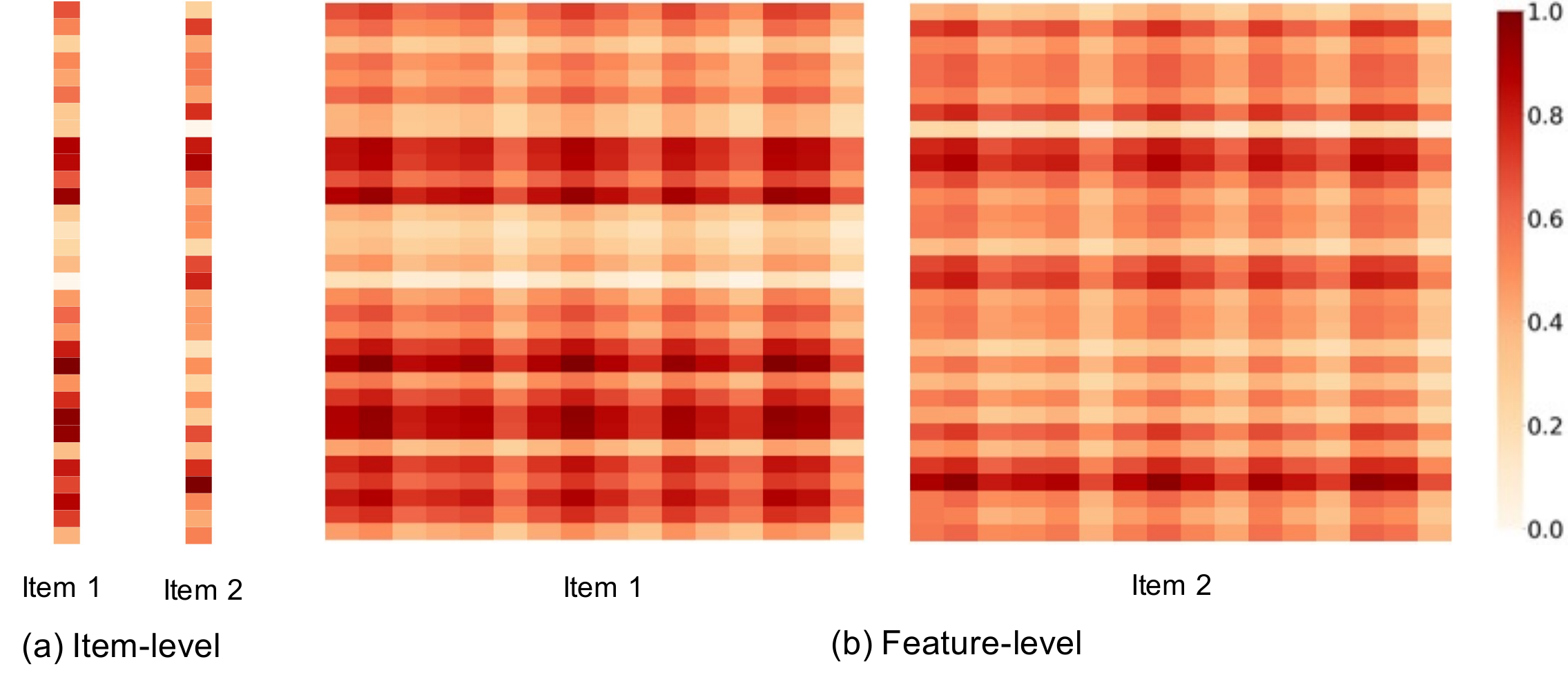}
    \caption{Visualization of the item- and feature-level attention for predicting a user preference to two items.}
    \label{figrq5}
\end{figure}

In the proposed feature-level attention ICF method, we claim that for recommending a target item to a user,  the user's historically interacted items contribute differently to the prediction (i.e., item-level attention). Furthermore, the features of those historical items are not equally important to the target item for prediction (i.e., feature-level attention). In this section, we would like to  visualize the attention weights of items and features to two different target items to validate our viewpoint. As shown in Fig.~\ref{figrq5}, we use heatmap to visualize: 1)  the attentions of historical items to the target item; and 2)  the  attentions of historical items' features to the target item. The color scale represents the intensities of attention weights, where a deeper color indicates a higher value and a lighter color indicates a lower value.

\textbf{Item-level attention.} Fig.6 (a) visualizes the attention weights of 32 items that were consumed by the user (in the ML-1m dataset) with respect to two target items. The y-axis represents the dimensions of attention vector (i.e., 32). At the beginning, the user attention on all aspects (or dimensions) is set to be the same (i.e., the all the values in the
attention vector is 1). During the training process, the attention
weight of each dimension will gradually adjust and finally converge to a constant value. Fig.6 (a) visualizes the converged values for two target items (i.e., item 1 and item 2). We can see that the attention weights for 64 historical items are not the same for the two items,  demonstrating that the historical items contribute differently to different target items.

\textbf{Feature-level attention.} Fig.6 (b) illustrates the attention weights of different features of 32  historical items towards two different items (in the ML-1m dataset). Note that the embedding size we used is 16, therefore there are 16 features in the figure.  The x-axis represents the dimensions of attention vector (i.e., 16), and y-axis represents different items that the user historically interacted. We can see that: 1) The attentive weights of all the features of some items are relatively larger than those of other items. As we discussed in Section 3.2, this is because some items are relevant to the target item while others are irrelevant. This also indicates the importance of the item-level attention. 2) For the same target item, the attention weights of different features are different for those historical items. And 3) for different target items, the attention weights are also different for the same historical item. This observation can well support our assumption for designing the feature-level attention: historical items contribute differently on the feature-level for recommendation in ICF models.

% \begin{table}[t]
%     \caption{Calculation results (\%) of DeepICF and FAMR when the embedding size is 16.}\smallskip
%     \centering
%     \resizebox{.99\textwidth}{!}{
%         \begin{tabular}{|c|c|c|c|c|c|c|c|c|c|c|c|c|c|c|c|c|}
%             \hline
%             \multirow{2}*{\textbf{Methods}} & \multicolumn{2}{c|}{\textbf{Patio}} & \multicolumn{2}{c|}{\textbf{Music}} & \multicolumn{2}{c|}{\textbf{Grocery}} & \multicolumn{2}{c|}{\textbf{Beauty}} & \multicolumn{2}{c|}{\textbf{Clothing}} & \multicolumn{2}{c|}{\textbf{Home}}\\
%             \cline{2-13}
%             ~ & \textbf{HR} & \textbf{NDCG} & \textbf{HR} & \textbf{NDCG} & \textbf{HR} & \textbf{NDCG} & \textbf{HR} & \textbf{NDCG} & \textbf{HR} & \textbf{NDCG} & \textbf{HR} & \textbf{NDCG}\\
%             \hline\hline
%             DeepICF & 26.57 & 12.42 & 60.96 & 37.31 & 50.94 & \textbf{31.26} & 49.76 & 30.85 & 43.24 & \textbf{26.56} & 48.98 & 29.83\\
%             \hline
%             FAMR & \textbf{33.51} & \textbf{17.23} & \textbf{63.06} & \textbf{38.87} & \textbf{52.37} & 30.95 & \textbf{52.48} & \textbf{32.95} & \textbf{43.64} & 25.88 & \textbf{50.43} & \textbf{30.66}\\
%             \hline
%         \end{tabular}
%     }
%     \label{tab5}
% \end{table} 

\section{Conclusion} \label{sec6}
In this work, we advocate the importance of modeling user diverse intents to items in recommendation and present a feature-level attention model for ICF models. The proposed model distinguishes the contributions of different features of a historical item to the target item for prediction. In this way, our model  captures user intents at the feature-level of item embeddings. In addition, we design a light attention neural network to combine the item- and feature-level attentions for neural ICF models. It is model-agnostic and easy-to-implement in ICF models. To show its effectiveness, we apply it to the recently proposed NAIS and DeepICF models and evaluate its effectiveness on six public datasets. The superior performance over several competitive baselines demonstrates the benefits of modeling the impact of different features (in item embeddings) for recommendation. 

We hope this work can shed light on modeling user preference at a fine-grained level to capture user diverse intents on adopting items for recommendation, and can motivate more researches in this direction in the future. Because it typically needs more data to model user preference on such a fine-grained level, an interesting future study is to exploit the rich side information, such as reviews and knowledge-graphs, in the modeling. In addition, how to leverage the fine-grained preference modeling to provide better interpretation for recommendations is also worth studying.

%%%%%%%%%%%%%%%%%%%%%%%%%%%%%%%%%%%%%%%%%%%%%%%%%%%%%%%%%%%%%%%%%%%%%%%%%%%%%%%%

\bibliographystyle{ACM-Reference-Format}
\bibliography{ref}

\end{document}